\documentclass[doublecol]{epl2}
\usepackage{amssymb}
\usepackage{amsbsy}
\usepackage{amsmath}

\newcommand{\vek}[1]{\bm{#1}}

\newcommand{\ip}[2]{\langle \! \langle #1;#2 \rangle \! \rangle}
\newcommand{\av}[1]{\langle #1 \rangle}

\title{Density Dynamics from Current Auto-Correlations at Finite Time- and Length-Scales}

\shorttitle{Density Dynamics from Current Auto-Correlations}

\author{Robin Steinigeweg\inst{1}\thanks{\tt{\email{rsteinig@uos.de}}} \and
        Hannu Wichterich\inst{1}\thanks{\tt{\email{hwichter@uos.de}}} \and
        Jochen Gemmer\inst{1}}

\shortauthor{R. Steinigeweg \etal}

\institute{\inst{1}
Fachbereich Physik,
Universit\"at Osnabr\"uck,
Barbarastrasse 7,
D-49069 Osnabr\"uck, Germany
}

\pacs{05.60.Gg}{Quantum transport}
\pacs{05.30.-d}{Quantum statistical mechanics}
\pacs{05.70.Ln}{Nonequilibrium and irreversible thermodynamics}

\abstract{
We consider the increase of the spatial variance of some
inhomogeneous, non-equilibrium density (particles, energy, etc.)
in a periodic quantum system of condensed matter-type. This is
done for a certain class of initial quantum states which is
supported by static linear response and typicality arguments. We
directly relate the broadening to some current auto-correlation
function at finite times. Our result is not limited to diffusive
behavior, however, in that case it yields a generalized Einstein
relation. These findings facilitate the approximation of diffusion
constants/conductivities on the basis of current auto-correlation
functions at finite times for finite systems. Pursuing this, we
quantitatively confirm the magnetization diffusion constant in a
spin chain which was recently found from non-equilibrium bath
scenarios.
}

\begin{document}

\maketitle

Any current of some physical quantity like, e.g., electrons
through solids is either induced  by an external mechanical force
$\vek{F}$ (e.g., electric field) or a  spatially non-uniform
density (density gradient $\vek{\nabla} \rho$). For systems
featuring normal transport the currents in those two cases are
routinely assumed to be determined by
\begin{equation}
\vek{j }= \kappa \, \vek{F} \, , \quad \vek{j} = -{\cal D} \,
\vek{\nabla} \rho \, , \label{twocases}
\end{equation}
where $\vek{j}$ denotes the current; $\kappa$ the conductivity;
and $\cal D$ the diffusion constant. As well-known, the derivation
of the l.h.s.~of Eq.~(\ref{twocases}) from linear response theory
is rather straightforward and leads to the Kubo formula
\cite{kubo1991} which nowadays is a standard approach to transport in
quantum systems \cite{zotos1999, heidrichmeisner2003, saito2003, 
zotos2003,benz2005, mejiamonasterio2005}. However, a direct derivation
of the r.h.s.~of Eq.~(\ref{twocases}) from the corresponding scenario,
i.e., without any external force appears to be more subtle. The overview
paper \cite{zwanzig1965} by Zwanzig, e.g., lists essentially six
families of approaches, each of them based on different sorts of
assumptions such as Fokker-Planck dynamics in the space of
relevant observables \cite{kirkwood1946, green1952, degroot1962},
Onsager's regression hypothesis \cite{onsager1931, kubo1957},
restriction of the dynamics to local equilibrium states
\cite{mori1958}, etc. These subtleties are especially disturbing
in the context of heat conduction, since there simply is no
practical external force which could cause a heat current
\cite{luttinger1964, garrido2001, saito2002, lepri2003, wu2008}.
However, all of the above approaches eventually give the diffusion
constant in terms of current auto-correlation functions also,
i.e., proportional to the conductivity. Of course, this is what
Einstein and Smolouchowski firstly suggested in their
ground-breaking work on Brownian motion. The above approaches are
comprehensively covered in textbooks like, e.g., \cite{kubo1991}.
Therein the interested reader may also find a discussion of their
consistency and implications. In order to motivate our present
alternative approach to this extensively debated subject, we
simply discuss here a standard formulation rather than going
through the above discussion. In \cite{kubo1991}, e.g., one finds
the following expression for a particle diffusion coefficient
\begin{equation}
{\cal D} = \frac{1}{k T \partial n / \partial \xi} \, \lim_{\omega
\rightarrow 0} \int \limits_0 ^{\infty} \text{d}t \, e^{i \omega
t} \lim_{\overset{\scriptstyle q \rightarrow 0}{\scriptstyle L
\rightarrow \infty}} \frac{\langle J'_{-q}(0); J'_{q}(t)
\rangle}{L^3} \, , \label{kubdiv}
\end{equation}
where $n$ is the equilibrium particle density; $\xi$ the chemical
potential; $J'$ denotes the ``random current''; $J'_q$ its Fourier
component with wavevector $q$; the brackets encode the Kubo-inner
product; and $L^3$ is the volume of the system. Conceptually, the
most challenging part is probably to concisely show that in the
above limit the auto-correlation of the random current may be
replaced by the auto-correlation of the ``true current'', i.e.,
$\langle J'_{-q}(0); J'_{q}(t) \rangle \rightarrow \langle
J_{-q}(0); J_{q}(t) \rangle$ (since the former involves the
``irrelevant part'' in  terms of projection methods), however,
this replacement appears to be generally accepted. Practically,
the limits $q \rightarrow 0$, $\omega \rightarrow 0$ imply that
only the infinitely slow dynamics of density structures of
infinite length scale can be expected to behave diffusively with
the above diffusion coefficient. So, even if one knew the exact
current auto-correlation function for the infinite system, that
would not bare implications on the dynamics at finite time- or
length-scales. And, even worth, if one has some information on the
current auto-correlation function but only up to a finite time,
(\ref{kubdiv}) does not allow for any conclusions on the dynamics.
Both issues become manifest, if the integral in (\ref{kubdiv})
does not converge, then no information results except for
the ``non-diffusiveness'' of the  dynamics. Furthermore, it may
be a little subtle to generalize (\ref{kubdiv}) to, e.g., a
micro canonical ensemble.

Thus, in this paper we approach the subject neither from
projection techniques nor from linear response. Instead we
(somewhat arbitrarily) ``coarse-grain'' the quantum system
spatially into subunits. The density profile and the current
operator are then discretely formulated on the basis of this
coarse-grained description. Afterwards a specific class of initial
states featuring such a non-uniform density profile is introduced.
Then, simply by applying Heisenberg's equation, the evolution of
the variance of the density profile is expressed in terms of a
double temporal integral of the current auto-correlation function.
Based on this result, we discuss the connection between diffusion
coefficient and conductivity. We further discuss the implications
of our specific choice for the initial state w.r.t.~external
perturbations and quantum typicality. Finally, we numerically
calculate the integral of the magnetization current
auto-correlation function in a XXZ spin chain and quantitatively
compare the outcome to recent results on the diffusion constants
for such a system from non-equilibrium bath scenarios.

The above mentioned periodic spatial coarse-graining scheme is
introduced to facilitate a consistent definition of a local
current. It is most conveniently explained for (but not limited
to) an one-dimensional system. To those ends the transported
quantity $\hat{X}$ as well as the Hamiltonian $\hat{H}$ are
decomposed into formally identical addends which correspond to
different positions, i.e., $\sum_{\mu} \hat{x}_\mu = \hat{X}$,
$\sum_{\mu} \hat{h}_\mu = \hat{H}$. Thus, $\hat{x}_\mu$ is a local
density of the transported quantity. Note that the $\hat{h}_\mu$
may be defined on or in between the positions of the
$\hat{x}_\mu$, or both. We consider quantities which are conserved
on the full system, i.e., $[\hat{H}, \hat{X}]=0$. We further
require
\begin{equation}
\dot{\hat{x}}_\mu = i [ \hat{H}, \hat{x}_\mu ] =
\underbrace{\imath [ \hat{h}_{\mu^-}, \hat{x}_\mu ]}_{\equiv
\hat{j}_{\mu-1}} + \underbrace{\imath [ \hat{h}_{\mu^+},
\hat{x}_\mu ]}_{\equiv -\hat{j}_{\mu}} \, . \label{heisloc}
\end{equation}
Here, $\hat{h}_{\mu^-}$ ($\hat{h}_{\mu^+}$) is supposed to denote
the local subunit of the Hamiltonian which is located directly on
the l.h.s.~(r.h.s.) of $\hat{x}_\mu$. This implies a kind of
locality. However, such a description may always be at least
approximately enforced, if the interactions are reasonably
short-ranged. It may require the usage of $\hat{h}_\mu$ that are
larger than a single elementary cell. Routinely, the comparison
with a continuity equation suggests a definition of local currents
according to the scheme indicated in Eq.~(\ref{heisloc})
\cite{zotos1999,heidrichmeisner2003,benz2005,gemmer2006}. This is
consistent, if $[ \hat{h}_{\mu^+}, \hat{x}_\mu ] + [
\hat{h}_{(\mu+1)^-}, \hat{x}_{(\mu+1)}] = 0$. The latter holds, if
$\hat{X}$ is globally conserved.

We now define the class of initial states we are going to
consider. Those read
\begin{equation}
\rho(0) \equiv \rho_\text{eq} + \sum_\mu
\frac{\delta_\mu}{\epsilon^2} \, \rho_\text{eq}^\frac{1}{2} \,
\hat{d}_\mu \, \rho_\text{eq}^\frac{1}{2}  \, , \, \hat{d}_\mu
\equiv \hat{x}_\mu - \av{\hat{x}_\mu} \, . \label{inst}
\end{equation}
Here, $\rho_\text{eq}$ is any stationary (equilibrium) state of
the full system, i.e., $[\hat{H}, \rho_\text{eq}] = 0$. The
brackets $\av{\ldots}$ denote full equilibrium averages, i.e.,
$\av{\hat{A}} \equiv \text{Tr} \{ \hat{A} \, \rho_\text{eq} \}$.
Let $\ip{\hat{A}}{ \hat{B}}$ indicate the inner product
$\ip{\hat{A}}{\hat{B}} \equiv \text{Tr} \{ \hat{A} \,
\rho_\text{eq}^\frac{1}{2} \, \hat{B} \,
\rho_\text{eq}^\frac{1}{2} \} $. We then define
\begin{equation}
c(t, \mu - \nu) \equiv \frac{1}{\epsilon^2}
\ip{\hat{d}_\mu(t)}{\hat{d}_\nu} \, , \; \epsilon^2 \equiv
\sum_{\mu} \ip{\hat{d}_\mu(t)}{\hat{d}_\nu} \label{spacecorr}
\end{equation}
which clarifies the $\epsilon^2$ from (\ref{inst}). $\epsilon^2$
does not depend on time, for periodic systems it further does not
depend on $\nu$. For interpretational reasons we note that in this
case $\epsilon^2$  may be rewritten as $\epsilon^2 = \ip{\hat{X} -
\av{X}}{\hat{X} - \av{\hat{X}}} / L = \av{(\hat{X} -
\av{\hat{X}})^2} / L$ (here we exploited $[ \hat{X}, \hat{H} ] =
0$), where now and in the following $L$ indicates the number of
subunits in the full system. $\epsilon^2$ hence quantifies the
equilibrium fluctuations of the transported quantity. Thus,
w.r.t.~(\ref{kubdiv}), this implies $\epsilon^2 = kT \, \partial
n/ \partial \xi$.

We denote the actual expectation value of the local deviation of
the transported quantity from full equilibrium by $d_\mu(t)$,
i.e.,
\begin{equation}
d_\mu(t) \equiv \text{Tr} \{ \hat{d}_\mu (t) \, \rho(0) \} \, .
\label{actdev}
\end{equation}
This way we may write
\begin{equation}
d_\mu(t) = \sum_{\nu} c(t,\mu - \nu) \, \delta_{\nu} \, , \,
\text{thus} \, \sum_{\mu} d_\mu(t) = \sum_{\mu} \delta_{\mu} \, .
\label{dd}
\end{equation}
We are going to analyze the spatial variance of those deviations
from equilibrium, while we require them to be normalized to one,
i.e., $\sum_{\mu} d_\mu(0) = 1$. The normalization is implemented
by a corresponding choice of the $\delta_{\mu}$. Then we may
directly, quantitatively compare to a discrete diffusion equation,
as outlined below. The above mentioned spatial variance $W^2(t)$
simply reads
\begin{equation}
W^2(t) \equiv \sum_\mu \mu^2 \, d_\mu(t) - \Big [ \sum_\mu \mu \,
d_\mu (t) \Big ]^2 \, . \label{var}
\end{equation}
If now, hypothetically, the dynamics of the $d_\mu(t)$ were
generated by a discrete diffusion equation of the form
\begin{equation}
\dot{d}_\mu(t) = {\cal D}(t) \, [ \, d_{\mu-1}(t) - 2 \, d_\mu(t)
+ d_{\mu+1}(t) \, ] \, , \label{disc}
\end{equation}
then the evolution of the variance would read
\begin{equation}
\dot{[W^2]}(t) \approx 2 \, {\cal D}(t) \label{scalvar}
\end{equation}
which holds, if the $d_\mu(t)$ vanish at the ends of a chain or
are reasonably concentrated at a sector of a ring. (Such a
concentration will be assumed throughout this paper.) Exploiting this,
we are able to deduce a diffusion constant ${\cal D}(t)$ from the
evolution of the variance. To those ends we rewrite the variance
using (\ref{dd}) which yields
\begin{eqnarray}
W^2(t) &=& \sum_\mu \mu^2 \, \delta_\mu + \sum_\eta \eta^2 \,
c(t,\eta) \nonumber \\
&-& \Big [ \sum_\mu \mu \, \delta_\mu \Big ]^2 - \Big [ \sum_\eta
\eta \, c(t,\eta) \Big ]^2 \, . \label{newvar}
\end{eqnarray}
 If the system features ``space inversion symmetry'', which we
assume in the following, the last term on the r.h.s.~of the above
Eq.~(\ref{newvar}) vanishes. In order to relate the evolution of
$W^2(t)$ to a current auto-correlation function, it turns out to
be helpful to consider its second derivative w.r.t time. According
to (\ref{newvar}), this reads
\begin{equation}
\ddot{[W^2]}(t) = \sum_\eta \eta^2 \, \ddot{c}(t,\eta) \, .
\label{secder}
\end{equation}
We may evaluate this using Heisenberg's equation:
\begin{equation}
\ddot{c}(t,\eta) = -\frac{1}{\epsilon^2} \, \ip{[ \hat{H}, [
\hat{H}, \hat{d}_\eta(t) ]]}{\hat{d}_0} \label{heis}
\end{equation}
which may be rewritten as
\begin{equation}
\ddot{c}(t,\eta) = -\frac{1}{\epsilon^2} \, \ip{\imath [\hat{H},
\hat{d}_\eta(t) ]}{\imath [ \hat{H}, \hat{d}_0]} \label{heiscor}
\end{equation}
or, exploiting (\ref{heisloc}) and (\ref{inst}), as
\begin{equation}
\ddot{c}(t,\eta) = -\frac{1}{\epsilon^2} \, \ip{(\hat{j}_{\eta -1}
- \hat{j}_{\eta})}{(\hat{j}_{-1}-\hat{j}_0)} \, . \label{curcor}
\end{equation}
Inserting the above Eq.~(\ref{curcor}) into (\ref{secder}), and
exploiting again that $\ip{\hat{j}_{\eta
+\zeta}(t)}{\hat{j}_{\eta}}$ does not depend on $\eta$ and
further vanishes for $\zeta \rightarrow L$, we obtain
\begin{equation}
\ddot{[W^2]}(t) = \frac{2}{L \, \epsilon^2} \,
\ip{\hat{J}(t)}{\hat{J}} \, . \label{dercor}
\end{equation}
Here, $\hat{J}$ denotes not a local but the total current in the
full system, i.e., $\hat{J} \equiv \sum_\mu \hat{j}_\mu$.
According to (\ref{scalvar}), the diffusion constant corresponds
to the first derivative of the variance which reads
\begin{equation}
\dot{[W^2]}(t)= \dot{[W^2]}(t = 0) + 2 \int_0^t \text{d}t' \,
\frac{1}{L \, \epsilon^2} \, \ip{\hat{J}(t')}{\hat{J}} \, .
\label{fdercor}
\end{equation}
We suggest here to assume that $\dot{[W^2]}(t = 0) = 0$. This
surely holds for $T \rightarrow \infty$, since in that limit
$W^2(t)$ is symmetric w.r.t.~time. However, irrespective of $T$,
according to typicality arguments, there are overwhelmingly more
(pure) states corresponding to higher $W^2$ compared to any lower
$W^2$ \cite{reimann2007}. Thus, it is not to be expected that any
state evolves towards lower $W^2$, forwards or backwards in time,
unless it has been deliberately constructed to do so. We hence
expect $W^2(t)$ to be essentially symmetric w.r.t.~time,
irrespective of $T$. Comparing to (\ref{scalvar}), we eventually
conclude for the diffusion constant
\begin{equation}
{\cal D}(t)= \int_0^t \text{d}t' \, \frac{1}{L \, \epsilon^2} \,
\ip{\hat{J}(t')}{\hat{J}} \, , \label{difcont}
\end{equation}
which is the first main result of this work.

In order to compare this to force-driven transport (as calculated
from the Kubo formula), we write out the integrand in (\ref{difcont})
explicitly finding
\begin{equation}
\frac{1}{L \, \epsilon^2} \, \ip{\hat{J}(t)}{\hat{J}} = \frac{1}{L
\, \epsilon^2} \sum_{m, n} (p_m \, p_n)^\frac{1}{2} ||J_{mn}||^2
\cos(\omega_{mn} \, t) \label{dfunk}
\end{equation}
with $p_m \equiv (\rho_\text{eq})_{mm}$, $J_{mn}$ as the matrix
elements of the respective operators in the energy eigenbasis and
with $\omega_{mn} = E_m - E_n$ as the difference of the energy
eigenvalues $E_m$, $E_n$. Force-driven transport (within the
linear regime) is routinely described by a response function that
relates the current density to the external force:
\begin{equation}
j(t) = \int_{-\infty}^t \text{d}t' \, \Phi(t') \, F(t') \, .
\label{lresp}
\end{equation}
For a canonical equilibrium the low-frequency part of $\Phi(t)$,
for which $\omega_{mn} \ll kT$ applies, may be written as
\begin{equation}
\label{lresplow} \Phi^\text{low}(t)= \frac{1}{L \, kT} \sum_{m, n}
(p_m \, p_n)^\frac{1}{2} ||J_{mn}||^2 \cos(\omega_{mn} \, t) \, .
\end{equation}
Thus, defining the response to slowly varying fields as $\sigma(t)
\equiv \int _0^t \text{d}t' \, \Phi(t')$, we get for times $t \gg
\hbar / kT$
\begin{equation}
{\cal D}(t) = \frac{kT}{\epsilon^2} \, \sigma(t) \label{einst}
\end{equation}
which is a somewhat generalized Einstein relation, since it
applies to all quantities (not only particles). Obviously, in the
case of diffusive transport the time dependencies of ${\cal D}(t)$
and $\sigma(t)$ are expected to vanish.

In the following we will discuss the choice of the initial
non-equilibrium state in Eq.~(\ref{inst}). This choice is
essentially supported by two arguments.

({\bf I}.)~Assume the system was exposed (before any transport
dynamics starts) to an, additionally weak, static ``potential''
such that the perturbed Hamiltonian reads $\hat{H}' = \hat{H} +
\sum_{\mu} v_{\mu} \, \hat{x}_{\mu}$.  Then the corresponding
canonical equilibrium state surely is given by $\rho_\text{eq}' =
\exp(-\hat{H}' / kT) / \text{Tr} \{ \exp(-\hat{H}' / kT) \}$.
(Note that this is of the same form as a so-called ``local
equilibrium state''). Now, $\rho_\text{eq}'$ may be calculated for
small $v_{\mu}$ using static linear response \cite{kubo1991}.
Doing so, one finds that ``low-frequency'' contributions of the
``non-homogeneous'' parts of $\rho_\text{eq}'$ and $\rho(0)$ are
proportional to each other, i.e.,
\begin{eqnarray}
&& (\rho_\text{eq}' - \rho_\text{eq})_{mn} \propto (\rho(0) -
\rho_\text{eq})_{mn} \nonumber \\
&& \text{for } \, |\omega_{mn}| \ll kT \, , \, v_\mu \propto
\delta_\mu \label{leq}
\end{eqnarray}
with $\rho_\text{eq}$ as the canonical state w.r.t.~to $\hat{H}$
alone. All the dynamics of $\rho_\text{eq}'$ and $\rho(0)$ under
$\hat{H}$ stem from their inhomogeneous parts. All dynamics for
times $t \gg \hbar / kT$ is controlled by their low-frequency
contributions. In other words: if the initial density profile is
induced by a previous static external potential (which is then
removed), the broadening of this density profile can be expected
to be of the same form as the broadening corresponding to the
hypothetical initial state $\rho(0)$, as discussed in detail
above. Our considerations therefore apply to this frequently
discussed type of initial state.

({\bf II}.)~A possibly even stronger motivation for the
consideration of the initial states in Eq.~(\ref{inst}) may be
given along the lines of typicality \cite{goldstein2006,
popescu2006, reimann2007}. Recently, it has been pointed out that
in quantum systems with large Hilbert space dimensions the
evolutions of observables of the type $\langle \psi(t) | \hat{A} |
\psi(t) \rangle$ do not depend crucially on the details of the
initial state $| \psi(0) \rangle$ \cite{bartsch2009}. Concretely,
this means that the overwhelming majority of a set of initial
states $| \psi_n(0) \rangle$ which all feature the same initial
expectation value $a$ of an observable $\hat{A}$, i.e.~$\langle
\psi_n(0) | \hat{A} | \psi_n(0) \rangle = a$, will approximately
yield the same evolution of this expectation value, namely
$\langle \psi_n(t) | \hat{A} | \psi_n(t) \rangle \approx \text{Tr}
\{ \hat{A} \, \rho^\text{HE}(t) \}$, where $\rho^\text{HE}(t)$
represents a Hilbert space ensemble and explicitly reads
$\rho^\text{HE}(0) = 1 + f \hat{A}$, where $f = f(a)$ is a
pertinent scalar function of $a$. Although not proven so far, it
is natural to suggest a generalization of this dynamical
typicality to many observables such that for an initial set
specified by $\langle \psi_n(0) | \hat{A}_m | \psi_n(0) \rangle =
a_m$ there exist typical evolutions for the expectation values of
the type $\langle \psi_n(t) | \hat{A}_m | \psi_n(t) \rangle
\approx \text{Tr} \{ \hat{A}_m \rho^\text{HE}(t)\}$ but now with a
$\rho^\text{HE}(0) = 1 + \sum_m f_m \, \hat{A}_m$, where the $f_m
= f_m(\{ a_l \})$ are functions of the $a_l$. If we consider for
the moment a micro canonical ensemble in (\ref{inst}), we have to
choose $\rho_\text{eq} = \hat{\Pi}_E / \text{Tr} \{ \hat{\Pi}_E
\}$ with $\hat{\Pi}_E$ being a projector projecting out the
subspace spanned by the energy eigenstates corresponding to
energies within a certain interval around $E$. With this choice
the initial state in (\ref{inst}) reads $\rho(0) = \hat{\Pi}_E (1
+ \sum_\mu \delta_\mu / \epsilon^2 \, \hat{d}_\mu) \hat{\Pi}_E /
\text{Tr} \{ \hat{\Pi}_E \}$ which essentially is the same state
as the latter $\rho^\text{HE}(0)$ with $\hat{A}_{\mu} =
\hat{d}_{\mu}$ within the respective subspace. Thus, even if the
true physical initial state may not be of the precise form in
(\ref{inst}), the initial state in (\ref{inst}) nevertheless
generates the typical evolution of the spatial density profile. Or
to rephrase, if some initial state features a given spatial
density profile and lives within a given energy region, the
dynamical broadening of the density profile will most likely be
described by (\ref{fdercor}), irrespective of the details of the
initial state.

\begin{figure}[htb]
\includegraphics[width=1.0\linewidth]{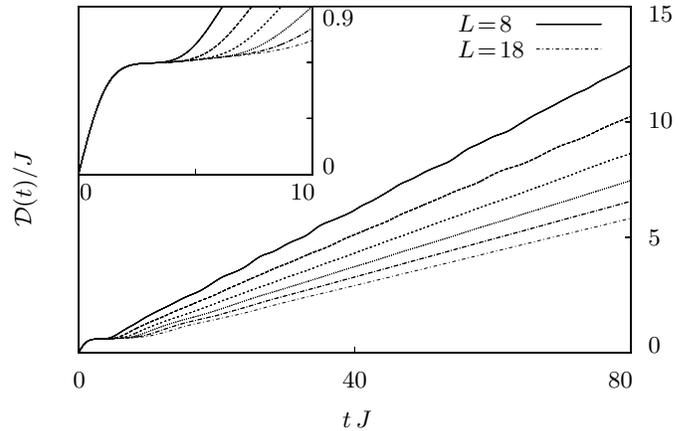} \caption{The diffusion
constant ${\cal D}(t)$, as defined by Eq.~(\ref{difcont}), for
magnetization transport (spin transport) in the anisotropic $s = 1/2$
Heisenberg chain (XXZ model) in the limit of high temperatures ($\beta
= 0$), evaluated numerically (exact diagonalization) for chain lengths
$L= 8$, $10$, $\ldots$, $18$. The curves for ${\cal D}(t)$ correspond
to the anisotropy parameter $\Delta = 1.5$ and are independent from the
concrete choice of the field strength $B$.}
\end{figure}

Finally, we compare approximations to the diffusion constant
${\cal D}(t)$, as given by Eq.~(\ref{difcont}), with recent
results in the literature. We approximate ${\cal D}(t)$ for finite
times, simply by numerically exact diagonalization of finite
systems. Concretely, we analyze magnetization transport,
i.e.~$\hat{x}_\mu = \hat{\sigma}^z_\mu / 2$, in the anisotropic $s
= 1/2$ Heisenberg chain (XXZ chain) with periodic boundary
conditions:
\begin{equation}
\hat{H} = \sum_{\mu = 1}^L \frac{J}{4} (\hat{\sigma}_\mu^x
\hat{\sigma}_{\mu+1}^x + \hat{\sigma}_\mu^y \hat{\sigma}_{\mu+1}^y
+ \Delta \, \hat{\sigma}_\mu^z \hat{\sigma}_{\mu+1}^z) +
\frac{B}{2} \hat{\sigma}_{\mu}^z \, . \label{spinr}
\end{equation}
Here, the operators $\hat{\sigma}_\mu^i$ ($i = x, y, z$) represent
the standard Pauli matrices (corresponding to site $\mu$); $J$
denotes the coupling strength; $\Delta$ is the anisotropy
parameter; and $B$ specifies the strength of the magnetic
field.

Fig.~1 exemplarily shows the result for $\Delta = 1.5$ and $\beta = 0$
for various lengths $L= 8$, $10$, $\ldots$, $18$. Note that in this
high temperature regime the  results are independent from $B$. At
short times (inset) we find an approximately constant ${\cal D}(t)
\approx {\cal D} = 0.6 \, J$. While the height of this ``plateau''
does not change with $L$, its width seems to increase gradually.
This increase is plausible, especially since the slope of
${\cal D}(t)$ at large $t$ coincides with the Drude weight and is
expected to vanish in the limit of $L \rightarrow \infty$
\cite{zotos1999,heidrichmeisner2003,benz2005}. Remarkably, ${\cal D}
= 2.4$ for $J = 4$ is very close to the value of $2.3$ which was found
in Ref.~\cite{prosen2008} from a numerically involved analysis of a
bath scenario. Furthermore, for a slightly different anisotropy
$\Delta = 1.6$ a similar Fig.~with ${\cal D} =0.55 \, J$ is found,
i.e., ${\cal D} = 0.022$ for $J = 0.04$. Also this result is in
very good agreement with the value of $0.0234$ in
Ref.~\cite{michel2008} which also addresses the high temperature
limit using bath scenarios.

These numerical findings support the
typicality statement, i.e., that the range of validity of
(\ref{difcont}) is not limited to the specific scenario, as
explicitly analyzed in the text. Thus, in a forthcoming paper
\cite{steinigeweg2009} by two of us Eq.~(\ref{difcont}) will be
applied to various spin models in detail.

\acknowledgments
We sincerely thank F.~Heidrich-Meisner for fruitful discussions.
Financial support by the Deutsche Forschungsgemeinschaft is
gratefully acknowledged.


\end{document}